\documentclass[twocolumn]{aastex63}

\graphicspath{{./}{figures/}}
\usepackage{pifont}
\usepackage{multirow}
\usepackage{amsmath}
\usepackage{CJK}
\usepackage{booktabs}
\usepackage{graphicx}
\usepackage{enumitem}

\shortauthors{Wang et al.}

\begin{document}

\title{A Channel to Form Fast-spinning Black Hole-Neutron Star Binary Mergers as Multimessenger Sources. II. Accretion-induced Spin-up}

\correspondingauthor{Ying Qin, Jin-Ping Zhu}
\email{yingqin2013@hotmail.com, jin-ping.zhu@monash.edu}

\author[0000-0001-9592-6671]{Zhen-Han-Tao Wang}
\affiliation{Guangxi Key Laboratory for Relativistic Astrophysics, School of Physical Science and Technology, Guangxi University, Nanning 530004, China}

\author[0000-0002-6442-7850]{Rui-Chong Hu}
\affiliation{Nevada Center for Astrophysics, University of Nevada, Las Vegas, NV 89154, USA}
\affiliation{Department of Physics and Astronomy, University of Nevada, Las Vegas, NV 89154, USA}

\author[0000-0002-2956-8367]{Ying Qin}
\affiliation{Department of Physics, Anhui Normal University, Wuhu, Anhui, 241002, China}

\author[0000-0002-9195-4904]{Jin-Ping Zhu}
\affil{School of Physics and Astronomy, Monash University, Clayton Victoria 3800, Australia}
\affiliation{OzGrav: The ARC Centre of Excellence for Gravitational Wave Discovery, Australia}

\author[0000-0002-9725-2524]{Bing Zhang}
\affiliation{Nevada Center for Astrophysics, University of Nevada, Las Vegas, NV 89154, USA}
\affiliation{Department of Physics and Astronomy, University of Nevada, Las Vegas, NV 89154, USA}

\author[0000-0003-0672-5646]{Shuang-Xi Yi}
\affiliation{School of Physics and Physical Engineering, Qufu Normal University, Qufu, Shandong 273165, China}

\author[0000-0001-7471-8451]{Qin-Wen Tang}
\affiliation{Department of Physics, School of Physics and Materials Science, Nanchang University, Nanchang 330031, China}

\author[0000-0002-7020-4290]{Xin-Wen Shu}
\affiliation{Department of Physics, Anhui Normal University, Wuhu, Anhui, 241002, China}

\author[0000-0002-6072-3329]{Fen Lyu}
\affiliation{Institute of Astronomy and Astrophysics, Anqing Normal University, Anqing 246133, People’s Republic of China}
\affiliation{Guangxi Key Laboratory for Relativistic Astrophysics, School of Physical Science and Technology, Guangxi University, Nanning 530004, China}

\author[0000-0002-7044-733X]{En-Wei Liang}
\affiliation{Guangxi Key Laboratory for Relativistic Astrophysics, School of Physical Science and Technology, Guangxi University, Nanning 530004, China}

\begin{abstract}
In this work, we investigate an alternative channel for the formation of fast-spinning black hole-neutron star (BHNS) binaries, in which super-Eddington accretion is expected to occur in accreting BHs during the stable mass transfer phase within BH-stripped
helium (BH--He-rich) star binary systems. We evolve intensive \texttt{MESA} grids of close-orbit BH--He-rich star systems to systematically explore the projected aligned spins of BHs in BHNS binaries, as well as the impact of different accretion limits on the tidal disruption probability and electromagnetic (EM) signature of BHNS mergers. Most of the BHs in BHNS mergers cannot be effectively spun up through accretion, if the accretion rate is limited to $\lesssim10\,\dot{M}_{\rm Edd}$, where $\dot{M}_{\rm Edd}$ is the standard Eddington accretion limit. In order to reach high spins (e.g., $\chi_{\rm BH} \gtrsim 0.5$), the BHs are required to be born less massive (e.g., $\lesssim3.0\,M_\odot$) in binary systems with initial periods of $\lesssim0.2-0.3\,{\rm days}$ and accrete material at $\sim100\,\dot{M}_{\rm Edd}$. However, even under this high accretion limit, $\gtrsim6\,M_\odot$ BHs are typically challenging to significantly spin up and generate detectable associated EM signals. Our population simulations suggest that different accretion limits have a slight impact on the ratio of tidal disruption events. However, as the accretion limit increases, the EM counterparts from the cosmological BHNS population can become bright overall.
\end{abstract}


\section{Introduction} \label{sec:intro}

Mergers of black hole--neutron star (BHNS) are the prime targeted gravitational-wave (GW) sources for the Advanced LIGO \citep{aasi2015}, Advanced Virgo \citep{acernese2015}, and KAGRA \citep{aso2013} GW detectors. At the end of the third observing run (O3), the LIGO-Virgo-KAGRA (LVK) Collaboration for the first time identified two high-confidence GWs from BHNS mergers, i.e., GW200105 and GW200115 \cite[][]{Abbott2021BHNS,nitz2021}. GW200105 was inferred to be a merger between a $8.9^{+1.2}_{-1.5}\,M_\odot$ BH with a near-zero spin of $\chi_{\rm BH} = 0.08^{+0.22}_{-0.08}$ and a $1.9^{+0.3}_{-0.2}\,M_\odot$ NS. The source of GW200115 has relatively lower component masses with the BH mass of $5.7^{+1.8}_{-2.1}\,M_\odot$ and the NS mass of $1.5^{+0.7}_{-0.3}\,M_\odot$. The BH component of GW200115 was reported to have a non-negligible mis-aligned spin with $\chi_{\rm BH} = 0.33^{+0.48}_{-0.29}$ \citep[see also][]{zhuxj2021}. However, the posterior distributions of component masses and BH spin of GW200115 could be highly impacted by the prior inference. Applying an astrophysically-motivated spin prior inference ($\chi_{\rm BH} \sim 0$) to GW200115 leads to much tighter constraints on the component masses, which were measured to be $7.0^{+0.4}_{-0.4}\,M_\odot$ and $1.25^{+0.09}_{-0.04}\,M_\odot$ \citep{mandel2021GW200115}, respectively. Furthermore, a few marginal events detected in the O3, e.g., GW190426\_1522155 and GW191219\_163120, could potentially originate from BHNS mergers if they have astrophysical origins \citep{abbott2021gwtc2,abbott2021gwtc21,abbott2023gwtc3}. As the LVK Collaboration began its latest campaign of O4, two BHNS candidates, S230518h and S230529a, have been reported recently \citep{ligo2023S230518h,ligo2023S230529ay}. More and more BHNS mergers are expected to be observed in the O4 and O5 observing runs, as well as the next-generation ground-based detectors, e.g., the Einstein Telescope \citep{Punturo2010} and the Cosmic Explorer \citep{Reitze2019}.

BHNS mergers that undergo tidal disruption are believed to power electromagnetic (EM) counterparts, such as kilonovae \cite[e.g.,][]{li1998,metzger2010,kasen2013,kasen2017,kyutoku2013,kyutoku2015,kawaguchi2016,kawaguchi2020,barbieri2019,zhu2020,zhu2022Long,darbha2021,gompertz2023} and gamma-ray bursts \cite[e.g.,][]{paczynski1986,paczynski1991,eichler1989,narayan1992,zhang2018,gottlieb2023}. The tidal disruption, however, is only expected for a fraction of the BHNS population under certain system parameters. More specifically, the systems that have a low-mass NS component with a stiff equation of state (EoS) and a low-mass BH with a high projected aligned spin are more easily to generate tidal disruption and generate bright EM signals\footnote{Weak EM signals could also be potentially produced for plunging events if the BH component is charged or the NS is strongly magnetized \citep[e.g.,][]{Zhang2019,Dai2019,Chen2021,Sridhar2021,D'Orazio2022,Yuan2023,Most2023}.} \citep[e.g.,][]{shibata2009,kyutoku2015,foucart2018,barbieri2019,raaijmakers2021,zhu2021no,zhu2021kilonova,zhu2022population,Clarke2023,Sarin2023}. Among the above parameters, the BH aligned spin could be one of the most critical parameter that affects the tidal disruption and EM detection for the population of BHNS merger in the universe. It is widely accepted that the majority of BHNS binaries are formed through the classical isolated common envelope (CE) channel \citep[e.g.,][]{Phinney1991,Tutukov1993,Hurley2002,Kalogera2007,Belczynski2016,Ablimit2018,Kruckow2018,Mapelli2018,Neijssel2019,Bavera2020,Belczynski2020,drozda2020,Shao2021,broekgaarden2021impact,Jaime2021,Hu2022,Mandel2022,Xing2023}. In this standard formation scenario, BHs typically formed at wide orbits are expected to possess low spins. The low BH spins in GW200105 and GW200115 revealed by the GW observations support that these BHNS binaries were originated from this standard formation scenario \citep{broekgaarden2021formation,zhu2022population,chattopadhyay2022,Biscoveanu2023,jiang2023}. Therefore, GW200105 and GW200115 could be likely plunging events and be hard to generate bright EM counterparts, which could explain why EM signals were not found by the follow-up observations of these GW events \citep[e.g.,][]{anand2021,zhu2021no,gompertz2021,D'Orazio2022}. The population synthesis simulations and analyses for population properties of O3 NSBH events \citep[e.g.,][]{zhu2021kilonova,zhu2022population,fragione2021NSBH,drozda2020,Biscoveanu2023,colombo2023multimessenger} also predicted that most of BHNS mergers in the universe could be plunging events without making any bright EM counterparts since these binaries unusually contain low-spin BH components, while detectable EM counterparts tend to be observed in BHNS systems with high aligned spin BHs. 

In order to possess high-spin BHs, the BHs or BH progenitors should be spun up before merging. One feasible scenario is that massive He-rich stars in close binaries after the CE phase can be efficiently spun up by strong tides to form fast-spinning BHs \citep{Qin2018,Bavera2020,Hu2022}, namely ``tidal-induced spin-up". Different groups have also independently investigated the spin-up of BHs in close binary systems through tidal interactions, i.e., \cite{Detmers2008,Kushnir2016,Olejak2021}. In order to form fast-spinning BHs in BHNS binaries, the NS could be born first due to a reversal in the mass ratio of progenitor stars during the Case A mass transfer (MT) stage. The massive He-rich star of the BH progenitor can be formed in close binaries later, which could be tidally spun up by the primary NS to potentially form a fast-spinning BH. In\defcitealias{Hu2022}{Paper I}\citetalias{Hu2022} \citep{Hu2022}, we carried out detailed binary modeling to investigate this peculiar channel, mainly focusing on the properties of resulting BHs and their associated tidal disruption probabilities \citep[see also][]{Jaime2021,Chattopadhyay2021}. Alternatively, the immediate progenitor of the BHNS binary could be a BH--He-rich star system formed after the CE phase or through chemically homogeneous evolution \citep{Marchant2016,Mandel2016,DeMink2016,Song2016,duBuisson2020,Riley2021,Zevin2021,Bavera2022a,Bavera2022b,Qin2022}. The BH can be spun up via the accretion during the stable MT phase \citep[see recent studies,][]{Qin2022RAA,shao2022,Zevin2022,Steinle2023}, called as ``accretion-induced spin-up". \cite{Xing2023} recently adopted a new binary population synthesis code \texttt{POSYDON} \citep{Tassos2023} to investigate the formation of merging BHNS binaries at solar metallicity, finding that the BH always forms first with a low spin magnitude (e.g. $\chi_{\rm BH} \lesssim 0.2$), and can subsequently reach moderately spin magnitudes (e.g. $\chi_{\rm BH} \lesssim 0.4$) through stable MT. It was found that high spins of BHs can be achieved if super-Eddington accretion is allowed \cite[e.g., ][]{Zevin2022,shao2022,Qin2022RAA}. Besides, to significantly spin up BHs in BH--He-rich star binaries, super-Eddington accretion must be at play \citep[see recent investigations in ][]{Steinle2023}, as the accretion time is very limited due to the short lifetime of massive He-rich stars. As pointed out by \cite{Steinle2023}, super-Eddington accretion is not impossible in principle, since the Eddington limit is dependent on the geometry of the accretion and various kinds of instabilities. It has been suggested that the majority of ultraluminous X-ray sources are stellar-mass BHs or NSs undergoing super-Eddington accretion \cite[e.g.,][]{Bachetti2013,Sutton2013,Walton2014,Middleton2015,Kitaki2017}. In particular, the Galactic X-ray binary SS 433 has been considered to be a super-Eddington accretion system (see the detailed review of theoretical and observational progress in \citealp{Fabrika2004,Okuda2009} and a recent update in \citealp{Cherepashchuk2020}). 

In this study, we utilize stellar structure and binary evolution code \texttt{MESA} \citep{Paxton2011,Paxton2013,Paxton2015,Paxton2018,Paxton2019,Paxton2023} to investigate how fast-spinning BHs in BHNS binaries as potential GW sources can be formed through super-Eddington accretion and their associated tidal disruption possibility. We first describe the setup of our binary evolution models in Section \ref{sect2}. Our main results of detailed binary evolution calculations and population simulations are presented in Section \ref{sect3}.  Finally, we summarize the conclusions with some discussion in Section \ref{sect4}.

\section{Methods} \label{sect2}

\subsection{Main Physics Adopted in \texttt{MESA}} \label{sec:MESAModel}
We adopt the released version r15140 of the \texttt{MESA} stellar evolution code \citep{Paxton2011,Paxton2013,Paxton2015,Paxton2018,Paxton2019,Paxton2023} to perform the detailed binary evolutionary calculations. Our single He-rich star models are built following the same method introduced in \cite{Qin2018,Bavera2020,Hu2022,Hu2023,Tassos2023,Qin2023,zhu2023mgap} with two steps: 1) create pure single He-rich stars at the zero-age helium main sequence (ZaHeMS); 2) relax the ZaHeMS models at given metallicity to reach the thermal equilibrium where the luminosity from core helium-burning just exceeds 99\% total luminosity. In our models, we evolve He-rich stars at the solar-metallicity environment with a metallicity of $Z=Z_{\odot}$, where $Z_{\odot}$ is set to be 0.0142 \citep{Asplund2009}. We model convection using the mixing-length theory \citep{MLT1958} with a mixing-length of $\alpha_{\rm mlt}=1.93$. Semiconvection \citep{Langer1983} with an efficiency parameter $\alpha_{\sc}=1.0$ is adopted in our modeling. We treat the boundaries of the convective zone using the Ledoux criterion and consider the step overshooting as an extension given by $\alpha_{\rm p} = 0.1 H_{\rm p}$, where $H_{\rm p}$ is the pressure scale height at the Ledoux boundary limit. The network \texttt{approx12.net} is used for our nucleosynthesis calculations. We treat rotational mixing and angular momentum transport as diffusive processes \citep{Heger2000}, including the effects of the Goldreich–Schubert–Fricke instability, Eddington–Sweet circulations, as well as secular and dynamical shear mixing. We adopt diffusive element mixing from these processes with an efficiency parameter of $f_{\rm c}=1/30$ \citep{Chaboyer1992,Heger2000}. 

We model stellar winds of He-rich stars as in \citetalias{Hu2022} and also take into account the rotationally-enhanced wind mass loss \citep{Heger1998,Langer1998} in the following:

    \begin{equation}\label{ml}
    \centering
    \dot{M}(\omega)= \dot{M}(0)\left(\frac{1}{1-\omega/\omega_{\rm crit}}\right)^\xi,
    \end{equation}
where $\omega$ and $\omega_{\rm crit}$ are the angular velocity and critical angular velocity at the surface, respectively. The critical angular velocity can be expressed as $\omega_{\rm crit} = \sqrt{(1- L/L_{\rm Edd})GM/R^3}$, where $L$, $M$, $R$ represent star's total luminosity, mass and radius, $L_{\rm Edd}$ is the Eddington luminosity, and $G$ is the gravitational constant. The default value of the exponent $\xi = 0.43$ is adopted from \citet{Langer1998}. For more details on wind mass loss enhanced by rotation, We refer readers to Section 5.2 in \cite{Qin2018}. No gravity darkening effect is accounted for \citep[see][for discussions on the impact of this process]{Maeder2000}. The mass transfer stability is also affected by the star's rotation. For instance, when an accretor (i.e., BH in this work) rotates, the accretion rate of the rotating accretor might be reduced by a factor of $(1-\omega/\omega_{\rm crit})$ \citep[e.g.,][]{Stancliffe2009,Ablimit2021,Ablimit2022,Oh2023}.

Stars have extended atmospheres and thus MT is expected to occur through the first Lagrangian point even when the star's radius $R_1$ is inside its Roche lobe $R_{\rm RL}$ \citep[namely Ritter scheme,][]{Ritter1988}. \cite{Kolb1990} further extended the Ritter scheme to cover the case $R_1 > R_{\rm RL}$. In this work, we modeled MT following the Kolb scheme \citep{Kolb1990} and the implicit MT method \citep{Paxton2015}. In addition, conservative mass transfer between a He-rich star and its companion is adopted for simplicity. For the tidal interactions, we apply the dynamical tides to He-rich stars with radiative envelopes \citep{Zahn1977}. The timescale of synchronization is calculated using the prescription mentioned in \cite{Zahn1977,Hut1981,Hurley2002}, while the tidal torque coefficient $E_2$ is adopted from the updated fitting formula as in \cite{Qin2018}.

We evolve He-rich stars until the carbon depletion is reached in the center. We assume a white dwarf (WD) is formed when the final carbon/oxygen (C/O) core mass is not higher than 1.37 $M_\odot$. In order to calculate the baryonic remnant mass, we adopt the ``\texttt{delayed}'' supernova (SN) prescription \citep{Fryer2012}. We also take into account the neutrino loss mechanism as described in the study by \cite{Zevin2020}. We then transform the baryonic remnant mass to the NS mass by Equation (13) and (14) described in \cite{Fryer2012}. In this work, the maximum NS mass is set to be 2.5\,$M_{\odot}$.

\subsection{Super-Eddington Accretion onto Stellar-mass BHs}

A BH accretes material through a disk at a rate $\dot{M}_{\rm acc}$ with the radiation efficiency $\eta$ determined by its innermost stable circular orbit, which is further dependent on the BH spin. It has been suggested that the natal spin of the BH could be typically negligible \cite[e.g.,][]{Fragos2015,Qin2018,Fuller2019,Belczynski2020}. For an initially non-rotating BH, its mass and BH dimensionless spin parameter increase through accretion \citep{Bardeen1970,Bardeen1972,Podsiadlowski2003} according to 

\begin{equation}
\label{equ:BHMass}
\dot{M}_{\rm BH} = (1- \eta) \dot{M}_{\rm acc},
\end{equation}
and

\begin{equation} \label{equ:BHSpin}
    \chi_{\rm BH} = \sqrt{\frac{2}{3}} \frac{M_{\rm BH,init}}{M_{\rm BH}}\left( 4 -\sqrt{18\left ( \frac{M_{\rm BH,init}}{M_{\rm BH}} \right )^2 - 2} \right),  
\end{equation}
for $M_{\rm BH} < \sqrt{6} M_{\rm BH,init}$, where the radiation efficiency is $\eta = 1 - \sqrt{1 - \left ( {M_{\rm BH}}/{3M_{\rm BH,init}} \right )^2}$ with the BH's initial mass $M_{\rm BH,init}$ and current mass $M_{\rm BH}$. 

Assuming an Eddington-limited accretion, i.e., the critical rate at which the outward force from the radiation pressure balances the inward gravitational pull, the maximum accretion rate is defined as 

\begin{equation}\label{m_edd}
\dot{M}_{\rm Edd} =  \frac{4\pi G M_{\rm BH}}{\kappa c \eta},
\end{equation}
where $c$ is the speed of light, and $\kappa$ is the opacity mainly determined by pure electron scattering, i.e., $\kappa=0.2(1+X)\,{\rm cm}^2\,{\rm g}^{-1}$ with the hydrogen mass fraction $X$.

\begin{figure}[tp]
     \centering
     \includegraphics[width=1\columnwidth, trim = 30 20 60 65, clip]{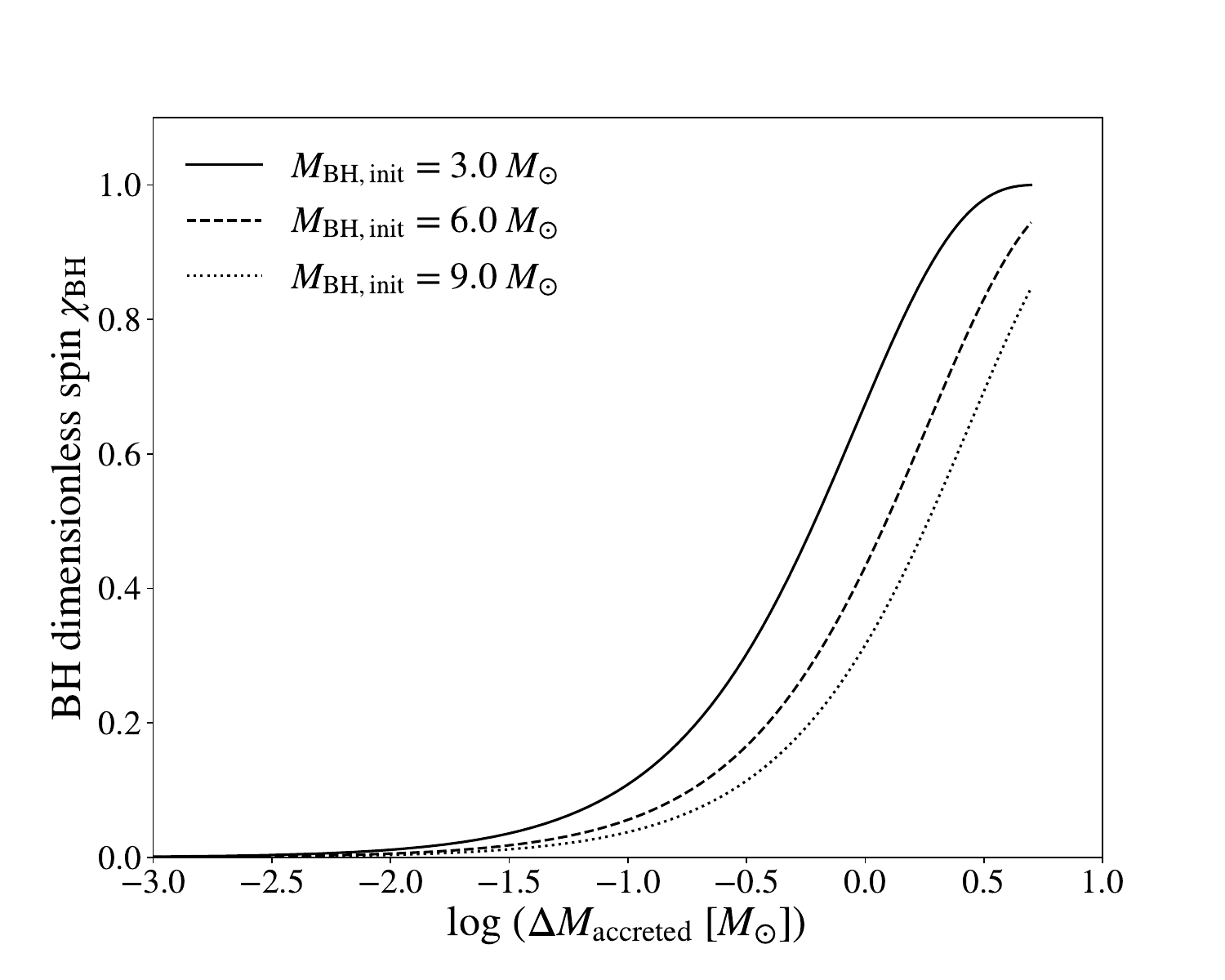}
     \caption{BH dimensionless spin $\chi_{\rm BH}$ as a function of its accreted material with different initial masses (Solid line: 3\,$M_\odot$; dashed line: 6\,$M_\odot$; dotted line: 9\,$M_\odot$).}
     \label{fig1}
\end{figure} 

As previously mentioned in the Introduction, super-Eddington accretion for the BH accretors is valid in principle. Here, we relax the standard Eddington-limited accretion to be 10\,$\Dot{M}_{\rm Edd}$ and 100\,$\Dot{M}_{\rm Edd}$ beyond which the BH can no longer accrete any material. By combining Equations (\ref{equ:BHMass}) and (\ref{equ:BHSpin}) to model the BH spin as a function of the accreted material, we show the evolution of the BH spin with its increased mass for different initial masses in Figure \ref{fig1}. We note that it is possible that the BH could spin down \cite[see Eqs. 42--44 in][]{Lei2017} when the accretion rate is high (e.g., exceeding Eddington accretion rate) due to launching a Blandford-Znajek jet \citep{Blandford1977}. Therefore, the spin magnitudes calculated in this work can be considered to be an upper limit.

\begin{figure} [tp]
     \centering
     \includegraphics[width=1\columnwidth, trim = 20 20 58 60, clip]{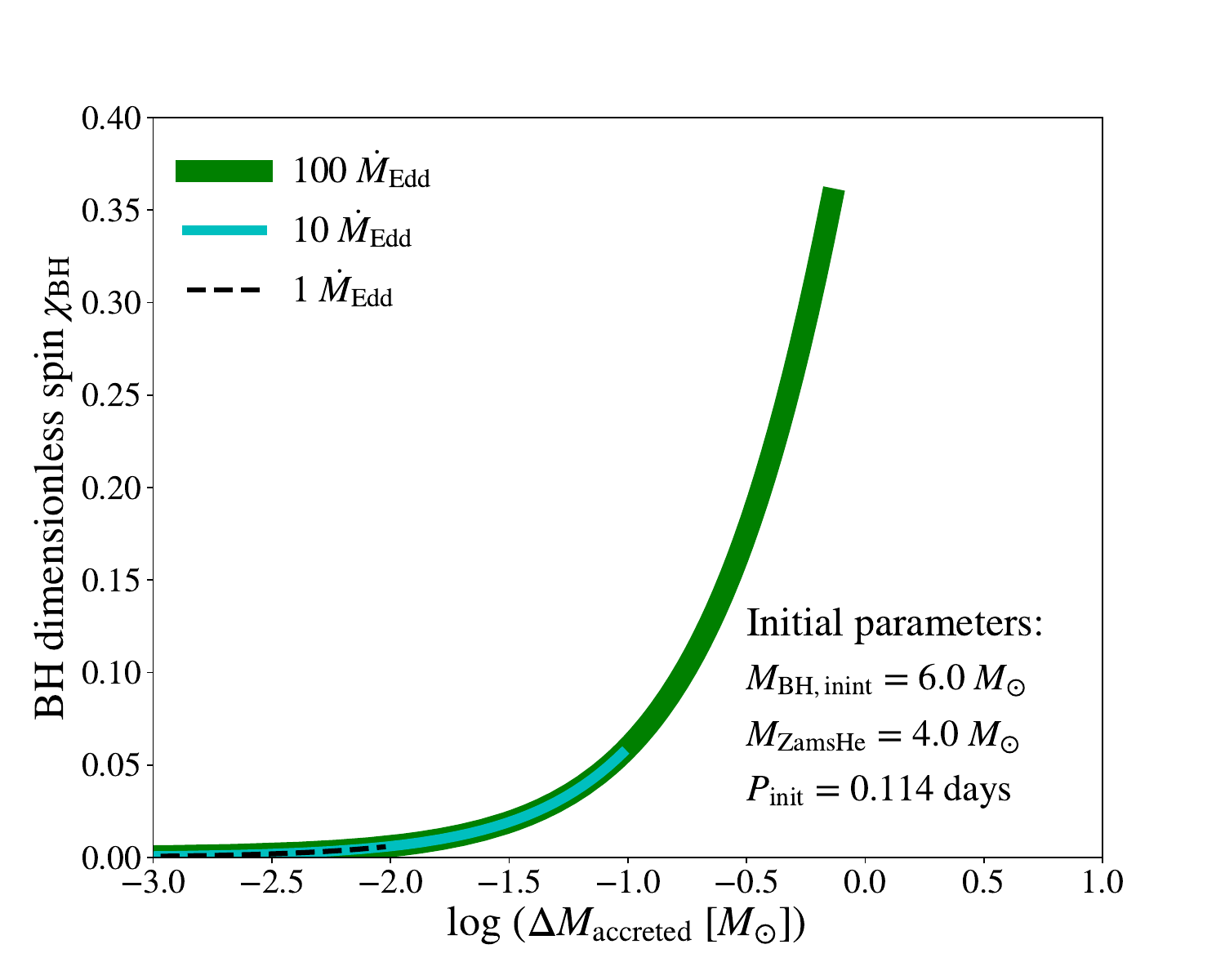}
     \caption{As a case study of our binary modeling, we show the BH dimensionless spin $\chi_{\rm BH}$ evolving with accreted mass for a binary system, which contains a $6\,M_\odot$ BH and a $4\,M_\odot$ He-rich star with an initial orbital period $P_{\rm init} = 0.114\,{\rm days}$. Three different accretion limits are considered: solid green line: $100\,\dot{M}_{\rm Edd}$, solid cyan line: $10\,\dot{M}_{\rm Edd}$; dashed black line: $1\,\dot{M}_{\rm Edd}$.}
     \label{fig2}
\end{figure}

As an example, we evolve a close binary of a $4\,M_{\odot}$ He-rich star and a $6\,M_{\odot}$ BH companion with an initial orbital period $P_{\rm init} = 0.114\,{\rm days}$. Three different accretion limits (1, 10, and $100\,\Dot{M}_{\rm Edd}$) are assumed in this modeling. In all cases, the binary system is found to experience MT from the shell-helium burning He-rich star (so-called case BB MT\footnote{Case BA(BB) MT: MT initiates during the core-helium (shell-helium) burning phase for He-rich stars; Case BC MT: MT initiates after shell-helium burning phase for He-rich stars.}) onto the BH. For the standard accretion limit (1\,$\Dot{M}_{\rm Edd}$) assumption, the BH has a spin magnitude $\chi_{\rm BH}$ below 0.01 due to a very small amount of mass accretion (see black dashed line in Figure \ref{fig2}). At 10 times Eddington limits (10\,$\Dot{M}_{\rm Edd}$), the BH can accrete mass close to 0.1\,$M_{\odot}$ and its spin magnitude $\chi_{\rm BH}$ is still low (e.g., $\chi_{\rm BH}<0.05$, see cyan solid line). In contrast, given 100\,$\Dot{M}_{\rm Edd}$, the BH can be efficiently spun up and reach a moderate value ($\chi_{\rm BH}\sim0.4$) by accreting $>0.7\,M_{\odot}$ (see green solid line).

\subsection{SN Kick Impacted onto NSs in BHNS Binaries} In general, NSs are expected to receive natal kicks when they are born via core-collapse supernovae because of asymmetric ejection of matter and possibly neutrinos. Therefore, the effect of kicks on the NS
itself and on the properties of the binary system (including the angular momentum evolution) is important for the formation of various NS binaries \cite[i.e.,][]{Belz1999,Mandel2021,Willcox2021,Ablimit2024}. When the second-born NS is formed, the SN gives a kick velocity to the NS, affecting the orbit of the binary system or potentially disrupting the system. Although the BH generally has a spin along the pre-SN orbital angular momentum axis, the SN kick to the newborn NS could result in a misalignment angle between the BH spin and system orbit \citep[e.g.,][]{Hills1983,Brandt1995,kalogera1996,Mandel2020recipe,zhuxj2021}. SN kick could also affect the tidal disruption probability and the amount of ejecta mass for BHNS mergers, since they are determined by the BH spin projected onto the orientation of orbital angular momentum (projected aligned spin hereafter), rather than the intrinsic BH spin. We thus take into account the SN kick imparted onto the final BHNS binary system according to the framework of \cite{kalogera1996,Wong2012,Callister2021}. 

Following the prescription outlined in \cite{Mandel2020recipe}, the mean natal kick for NSs can be expressed as

\begin{equation}
\label{equ:kickVelocity}
    \mu_{\rm kick} = v_{\rm NS}\frac{M_{\rm CO}-M_{\rm NS}}{M_{\rm NS}},
\end{equation}
where $v_{\rm NS}$ is the scaling prefactor for NS, and $M_{\rm CO}$ the carbon-oxygen core mass. We adopt a default value of $v_{\rm NS} = 400\,{\rm km}\,{\rm s}^{-1}$ suggested by \cite{Mandel2020recipe}. An additional component drawn from a Gaussian distribution with a standard deviation of $0.3\mu_{\rm kick}$ is considered. Thus, the kick velocity is $v_{\rm kick}\sim\mathcal{N}(\mu = \mu_{\rm kick},\sigma = 0.3\mu_{\rm kick})$.

The orbit of the binary system undergoes changes due to the formation of the NS and SN kick. We assume a random SN kick direction and, then, obtain the velocity component along each axis (i.e., $v_{{\rm orb},x}$, $v_{{\rm orb},y}$, and $v_{{\rm orb},z}$). Our calculations assume that the orbit in the pre-SN stage is circular, which means that the orbital separation $r$ is equivalent to the initial semimajor axis $a_{\rm i}$. The orbital separation $r$ remains unchanged under the assumption of an instantaneous SN explosion. We then update the semimajor axis $a_{\rm f}$ and eccentricity $e$ of the binary system after the SN kick following the approach outlined by \cite{kalogera1996}, i.e.,

\begin{equation}
\begin{split}
    a_{\rm f} = &G(M_{\rm NS}+M_{\rm BH})\left[\frac{2G(M_{\rm NS}+M_{\rm BH})}{a_{\rm i}}-v_{\rm kick}^2-\right. \\
    &\left.v_{\rm orb}^2-2v_{{\rm kick},y}v_{\rm orb}\right]^{-1}
\end{split}
\end{equation}
and

\begin{equation}
    1-e^2 = \frac{(v_{{\rm kick},y}^2+v_{{\rm kick},z}^2+v_{\rm orb}^2+2v_{{\rm kick},y}v_{\rm orb})a_{\rm i}^2}{G(M_{\rm NS}+M_{\rm BH})a_{\rm f}},
\end{equation}
where $v_{{\rm kick},i}$ represents the $i$-th component of the kick velocity $v_{\rm kick}$. Here, we adopt $v_{\rm orb}$, the pre-SN orbital velocity of the NS progenitor relative to its BH companion, given in \cite{Wong2012} as follows:

\begin{equation}
    v_{\rm orb}^2 = G(M_{\rm BH}+M_{\rm pre-SN})\left(\frac{2}{r}-\frac{1}{a_{\rm i}}\right),
\end{equation}
where $M_{\rm pre-SN}$ is the pre-SN mass of the NS progenitor.

Following the same assumption suggested in \cite{Callister2021}, the binary system is disrupted if

\begin{equation}
    \beta < \frac{1}{2}+\frac{v_{\rm kick}^2}{2v_{\rm orb}^2}+\frac{v_{{\rm kick,}y}v_{\rm orb}}{v_{\rm orb}^2},
\end{equation}
where $\beta = (M_{\rm NS}+M_{\rm BH})/(M_{\rm pre-SN}+M_{\rm BH})$ is defined in Equation (7) of \cite{kalogera1996}. For the binary systems that survive after the SN kicks, we then calculate the merger time via GW emission \citep{Peters1964} by

\begin{equation}
\label{equ:mergerTime}
    T_{\rm merge} = \frac{5}{256}\frac{c^5a_{\rm f}^4}{G^3(M_{\rm NS}+M_{\rm BH})^2m_{\rm r}}T(e),
\end{equation}
where $m_{\rm r} = M_{\rm NS}M_{\rm BH}/(M_{\rm NS}+M_{\rm BH})$ is the reduced mass of the binary and $T(e) = (1+0.27e^{10}+0.33e^{20}+0.2e^{1000})(1-e^2)^{7/2}$ is adopted following an accurate analytical fitting as in \cite{Mandel2021ecc}.

The angle $\theta$ between the pre-and post-SN orbital planes can be calculated by

\begin{equation} \label{theta}
    \cos\theta = \frac{v_{{\rm orb},y}+v_{\rm orb}}{\sqrt{(v_{{\rm kick},y} + v_{\rm orb})^2 + v_{{\rm kick},z}^2}}.
\end{equation}
Thus, the dimensionless projected aligned spin can be expressed as $\chi_{{\rm BH},z} = \chi_{\rm BH}\cos\theta$.

\section{Results} \label{sect3}

\subsection{Various Mass Transfer Phases in the Formation of BHNS Binaries} \label{sec:BHNSResult}

\begin{figure*}
     \centering
     \includegraphics[width=1.0\textwidth, trim = 25 88 30 65, clip]{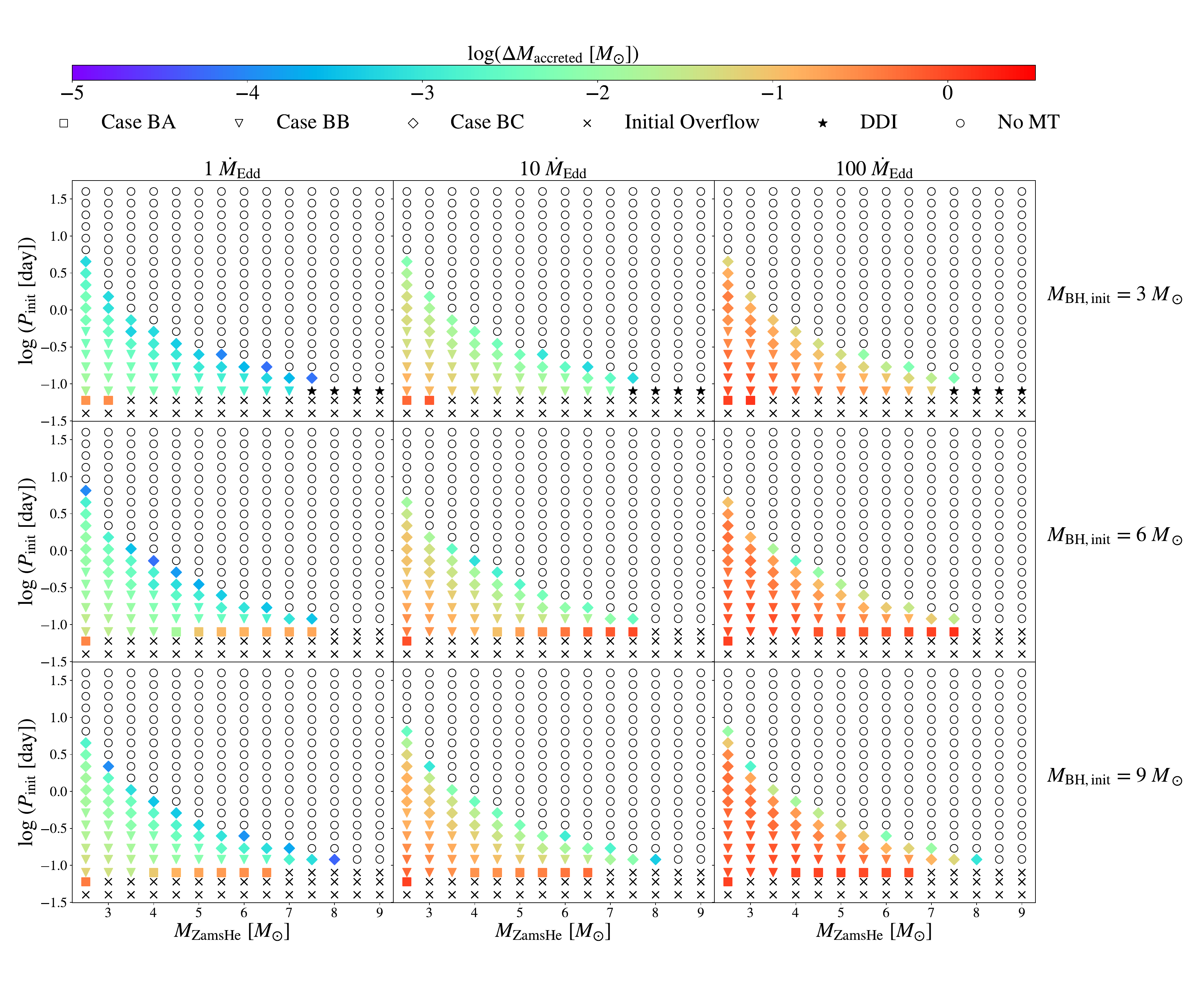}
         \caption{Accreted mass (see the color bar) as a function of the initial orbital period and initial mass of He-rich stars. Square: Case BA; triangle: Case BB; diamond: Case BC; cross: Initial Overflow; Star: DDI); circle: No MT. The \textit{three columns} correspond to different Eddington limits and the \textit{three rows} refer to different masses of the BH as a companion.}
     \label{fig3}
\end{figure*} 

As demonstrated above, BHs can be efficiently spun up under the condition of high accretion limits. Here we perform parameter space analysis of a close binary system consisting of a BH and a He-rich star. In particular, we investigate the impact of three key parameters including initial He-rich star mass, initial BH mass, and initial orbital period. We cover the initial He-rich star mass of $2.5-9.0\,M_{\odot}$ in steps of $0.5\,M_\odot$, BH as a point mass of $3.0-15.0\,M_{\odot}$ in steps of $3.0\,M_{\odot}$, and initial orbital period in a range of $0.04-40\,{\rm days}$ with $\Delta$log $(P_{\rm init}$[day]) $\sim$ 0.16 dex.

Let us begin by describing the upper-left panel of Figure \ref{fig3}. In the standard accretion assumption, the He-rich star's companion is assumed to be a BH with an initial mass of $3.0\,M_{\odot}$. Most binary systems are found to experience MT when the initial orbital period is shorter than $\sim1.0\,{\rm day}$. The more massive a He-rich star is, the shorter an orbital period is required for the binary system to undergo interactions in terms of all kinds of MT. This is because massive He-rich stars are more compact when compared with low-mass He-rich stars \cite[see their Figure 1 in][]{Qin2023}. Systems that have various interactions are divided into three categories (Case BA, Case BB, and Case BC) based on MT occurring in different evolutionary phases of He-rich stars. For binary systems with Case BA MT, the accreted mass onto a BH can reach $\sim0.3\,M_{\odot}$. For binary systems that experience Case BB MT, the accreted mass onto BHs can be higher than $0.01\,M_{\odot}$ when $M_{\rm ZamsHe} \leq 3.5 \, M_{\odot}$. Furthermore, BHs in all binary systems that have gone through Case BC MT can accrete mass $\sim$ $0.001\,M_{\odot}$. We also note that a small fraction of systems with higher initial mass and lower initial orbital periods (e.g., $P_{\rm init} \sim 0.1\,{\rm days}$ and $ 7.5\, M_{\odot} \leq M_{\rm ZamsHe} \leq 9.0 \, M_{\odot}$) are found to experience unstable MT, leading to the dynamical delayed instability \citep[DDI, see][]{Ivanova2003}. Additionally, the systems with an initial orbital period of $P_{\rm init} < 0.1\,{\rm day}$ undergo initial overflow through the first Lagrangian point, which is considered to be located in the forbidden parameter space in this work.

When considering $10\,\dot{M}_{\rm Edd}$ (see the top-middle panel in the first row), one can see that the initial conditions (initial mass and orbital period) of the binary systems play no impact on the parameter space of the MT. The stark difference, however, is that the system accretes more mass when compared with the standard Eddington limit. Such difference becomes even more obvious when considering $100\,\dot{M}_{\rm Edd}$ (see the top-right panel in the first row). For instance, the accreted mass onto a BH via Case BB (BC) MT can reach up to $\sim1.0\,M_{\odot}$ ($0.1\,M_\odot$). It is worth noting that the accreted mass through Case BA MT can even reach beyond $1.0\,M_{\odot}$. For more massive primary BHs ($M_{\rm BH,init} = 6.0\,M_{\odot}$ in the second row and $M_{\rm BH,init} = 9.0\,M_{\odot}$ in the third row), we note that more systems instead are found to undergo Case BA MT, producing much higher accreted mass even under the condition of the standard Eddington limit.

\begin{figure*}
     \centering
     \includegraphics[width=1.0\textwidth, trim = 25 88 30 65, clip]{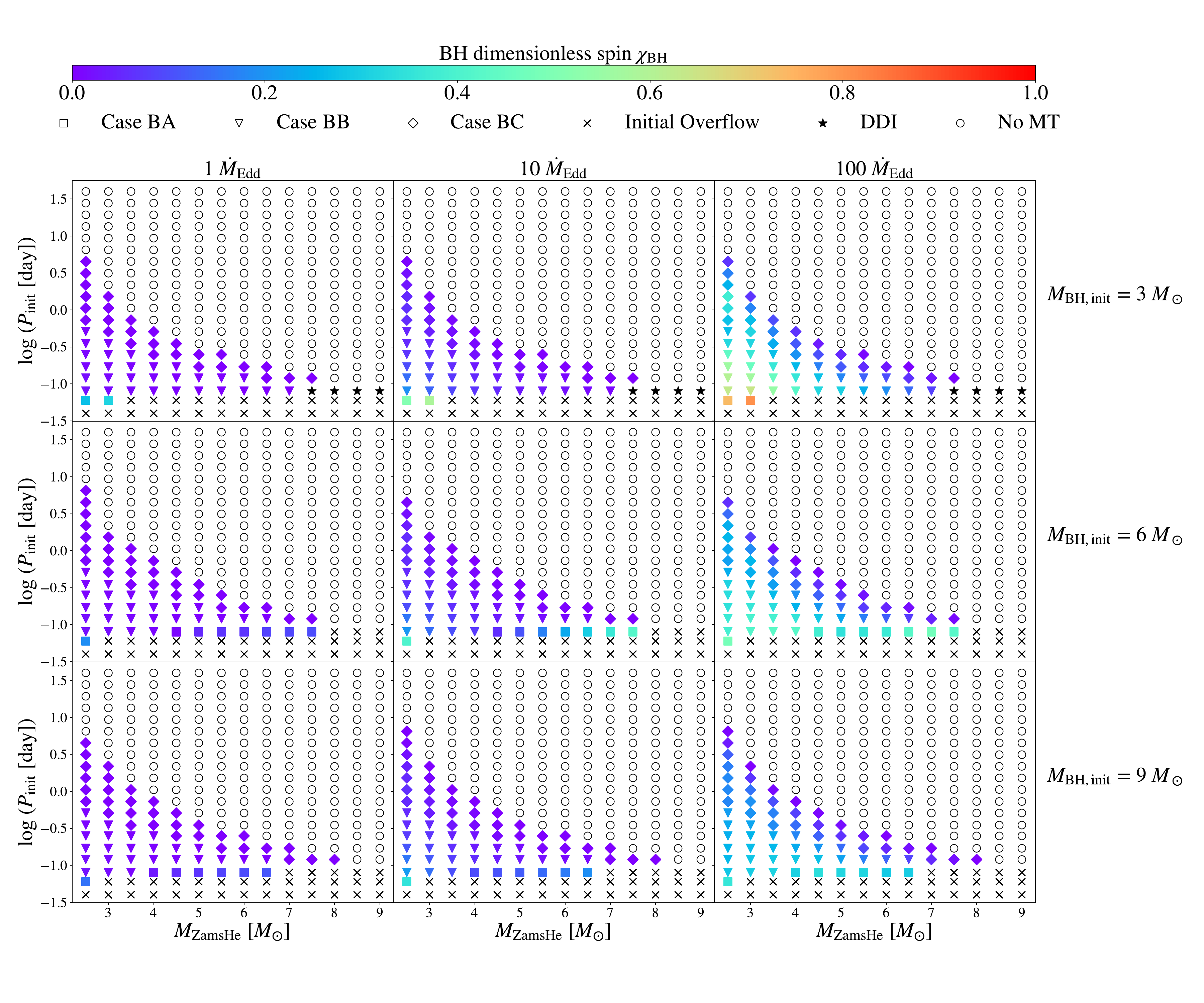}
     \caption{As in Figure \ref{fig3}, but the color bar corresponds to the final spin of the BH via accretion from its companion He-rich star.}
     \label{fig4}
\end{figure*} 

As presented in Figure \ref{fig2}, the Eddington limit plays a critical role in the accreted mass of the BH and thus its corresponding spin magnitude. We here then carry out a more detailed investigation of the impact of the binary's different initial conditions on the spin magnitudes. It is found in the upper-left panel of Figure \ref{fig4} that Case BB/BC MT cannot significantly spin up the BH (i.e., $\chi_{\rm BH}$  $\lesssim 0.1$) due to a limited time of accretion, while Case BA MT instead can allow the BH spin to reach a moderate value of $\chi_{\rm BH}\, \sim 0.3$. Similar results are obtained when a higher Eddington accretion limit (e.g. 10\,$\dot{M}_{\rm Edd}$, see the middle panel in the first row) is assumed. Notably, the BH can be efficiently spun up under 100\,$\dot{M}_{\rm Edd}$ assumption. The spin magnitude can reach a value of $\chi_{\rm BH}$ close to $\sim 0.7$ ($\sim 0.5$) for Case BB (Case BC) MT, even higher than 0.8 for Case BA MT. Nevertheless, the spin-up of a more massive BH companion becomes much less efficient when compared with a low-mass BH companion (see the second row for a 6\,$M_{\odot}$ BH and the third row for a 9\,$M_{\odot}$ BH, respectively), which has been demonstrated at the beginning of Section \ref{sect3} (see Figure \ref{fig1}). 

\begin{figure*}
     \centering
     \includegraphics[width=1.0\textwidth,trim = 25 0 222 0, clip]{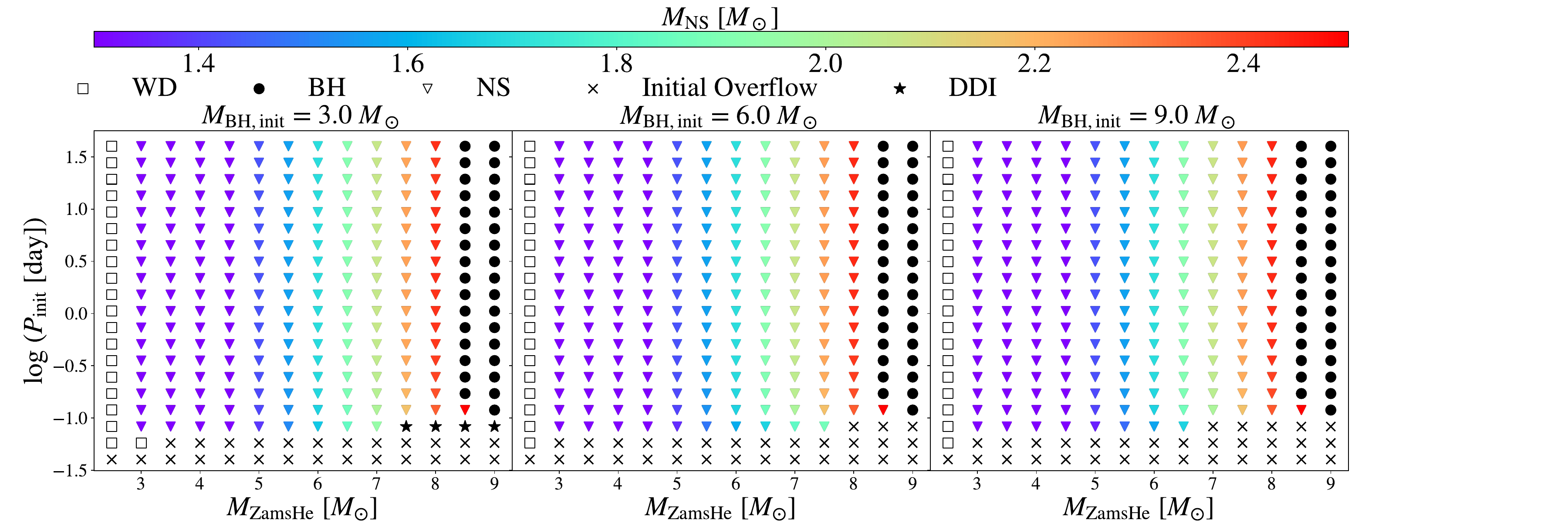}
     \includegraphics[width=1.0\textwidth,trim = 25 0 222 0, clip]{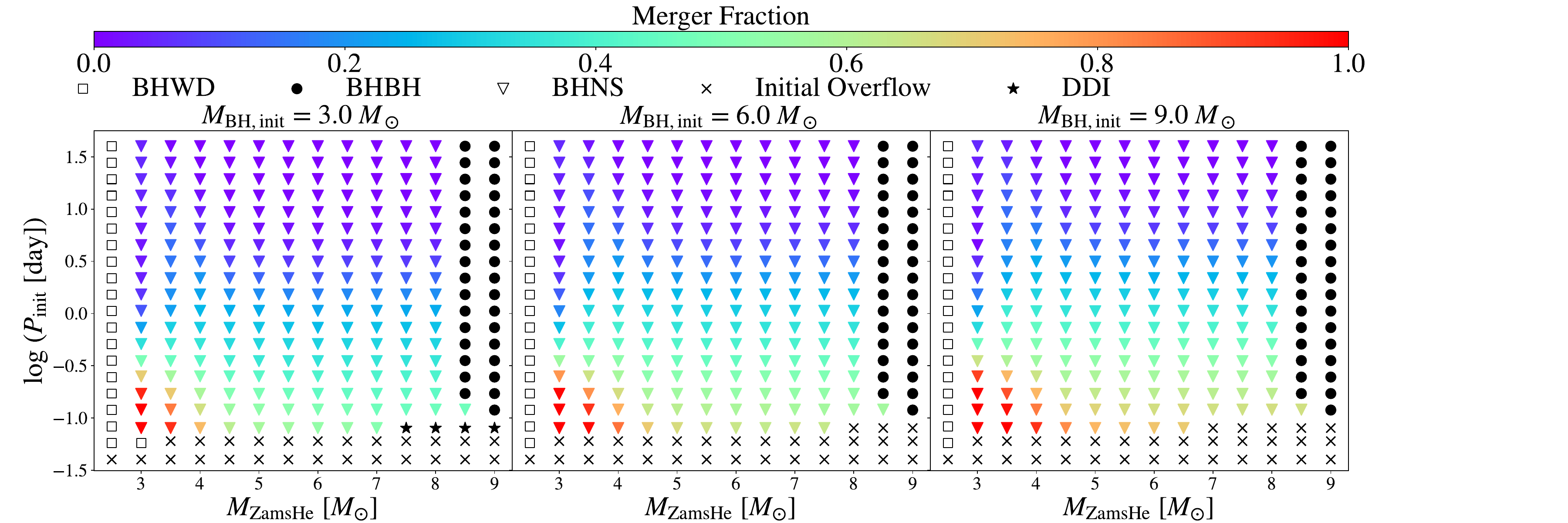}
     \includegraphics[width=1.0\textwidth,trim = 25 0 222 0, clip]{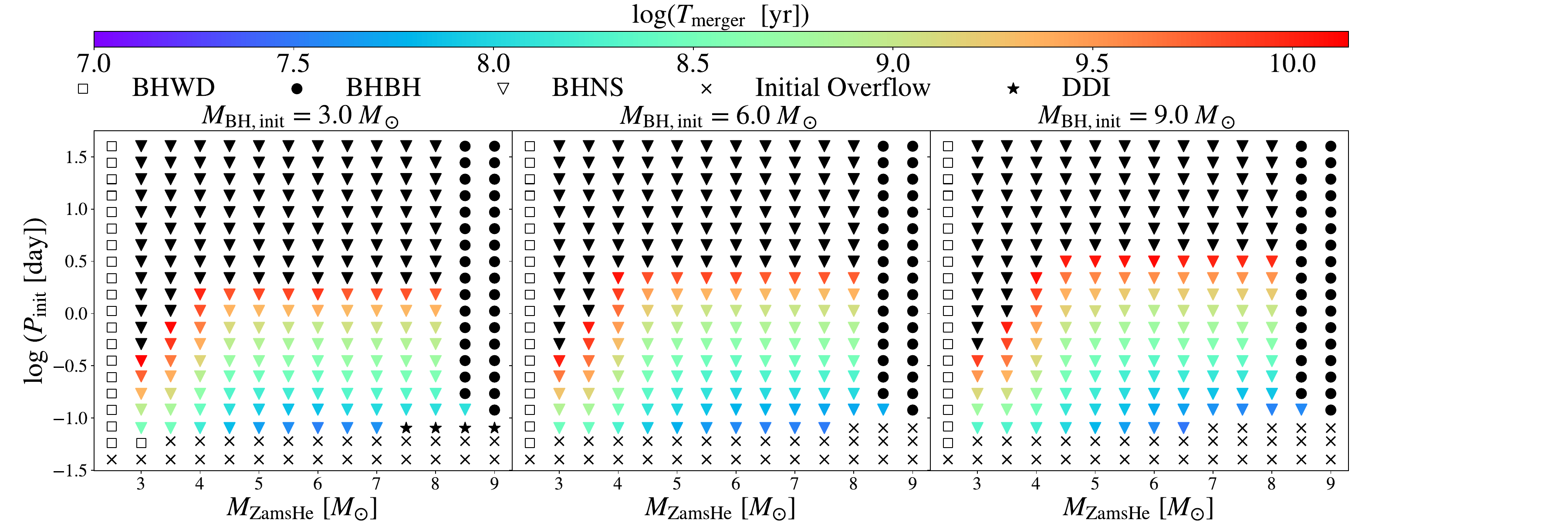}
     \caption{As in Figure \ref{fig3}, but three color bars from top to bottom rows refer to the mass of NS formed based on the ``\texttt{delayed}'' SN prescription \citep{Fryer2012}, the fraction of BHNS binaries that can be merged within Hubble time with SN kicks considered, and $T_{\rm merger}$ of BHNS binaries due to GW emissions, respectively. Square: WD; triangle: NS; dot: BH; cross: Initial Overflow; Star: DDI. \textit{Left panel}: 3.0\,$M_\odot$ BH companion; \textit{middle panel}: 6.0\,$M_\odot$ BH companion; \textit{right panel}: 9.0\,$M_\odot$ BH companion. For the merger fraction, we use black triangles to mark BHNS binaries whose merger times are beyond Hubble time. Different Eddington accretion limits have little impact on the mass of resulting NS, merger fraction, and $T_{\rm merger}$, and hence other similar figures corresponding to higher Eddington limits are not shown here.}
     \label{fig5}
\end{figure*} 
 
Figure \ref{fig5} presents various outcomes of the resulting compact binary objects (BHWD, BHNS, and BHBH), and only BHNS binaries are presented in this work. Since different accretion limits have little impact on the mass of newborn NS, merger fraction, and $T_{\rm merger}$, we only describe the results under the standard Eddington accretion assumption. In the left panel of the top row, we present the mass of NS following the ``\texttt{delayed}'' SN prescription \citep{Fryer2012} after the He-rich star reaches its carbon depletion in the center. Clearly, the initial orbital period has a slight impact on the final mass of the massive He-rich star and thus the resulting NS mass. This is because He-rich stars could be tidally locked by tides in close binaries \citep{Qin2018}, which allows them to have different rotation rates depending on their current orbital periods. Furthermore, rotation is found to enhance wind mass loss \citep[e.g.,][]{Langer1998}, which potentially reduces the resulting mass for massive stars \citep[see their Figure 3 in][]{lv2023}. It is also found that the BH companion with a slightly higher mass has negligible impact on the mass of the resulting NS (see the middle and right panel).

The SN kick imparted onto a newborn NS can change the separation of the binary systems, some of which might be disrupted. Similar to the same approach in \citetalias{Hu2022}, we obtain the kick velocity and kick direction by drawing $10^5$ times from a Gaussian distribution and a random distribution, respectively. As shown in the middle panels of Figure \ref{fig5}, the color represents the merger fraction of each BHNS system that can be merged within Hubble time. In the left panel of the middle row (i.e., the standard Eddington limit), we can see there are two main findings. First of all, the merger fraction of the binary system is higher when its initial orbital period is shorter. This is because more BHNS binaries are disrupted or too wide to merge within Hubble time. Besides, we note that the merger fraction decreases when increasing the initial mass of the He-rich star, resulting in a higher kick velocity by following the expression in Equation (\ref{equ:kickVelocity}). After the NS is formed, GW emission shrinks the orbit of the BHNS binaries by removing their orbital angular momentum, leading to the merger of the two compact objects. The timescale for the merger of a binary compact object due to GWs is calculated following Equation (\ref{equ:mergerTime}). In order to take into account the impact of SN kicks, we adopt the median of the merger time in $10^5$ times as $T_{\rm merger}$ in the bottom panels of Figure \ref{fig5}. It is expected that $T_{\rm merger}$ is shorter for BHNS binaries with an initially shorter orbit. Furthermore, $T_{\rm merger}$ is shorter when initial He-rich stars are more massive. This finding is mainly due to the anticorrelation between the total mass and the merging timescale \citep[see Section 6 for more details in][]{Qin2018}. Additionally, it is clearly shown that both the Eddington limit and the BH companion mass have a negligible impact on the merger fraction and $T_{\rm merger}$.

\subsection{Parameter Space of an NS Disrupted by a BH}

\begin{figure*}
     \centering
     \includegraphics[width=1.0\textwidth,trim = 25 0 222 0, clip]{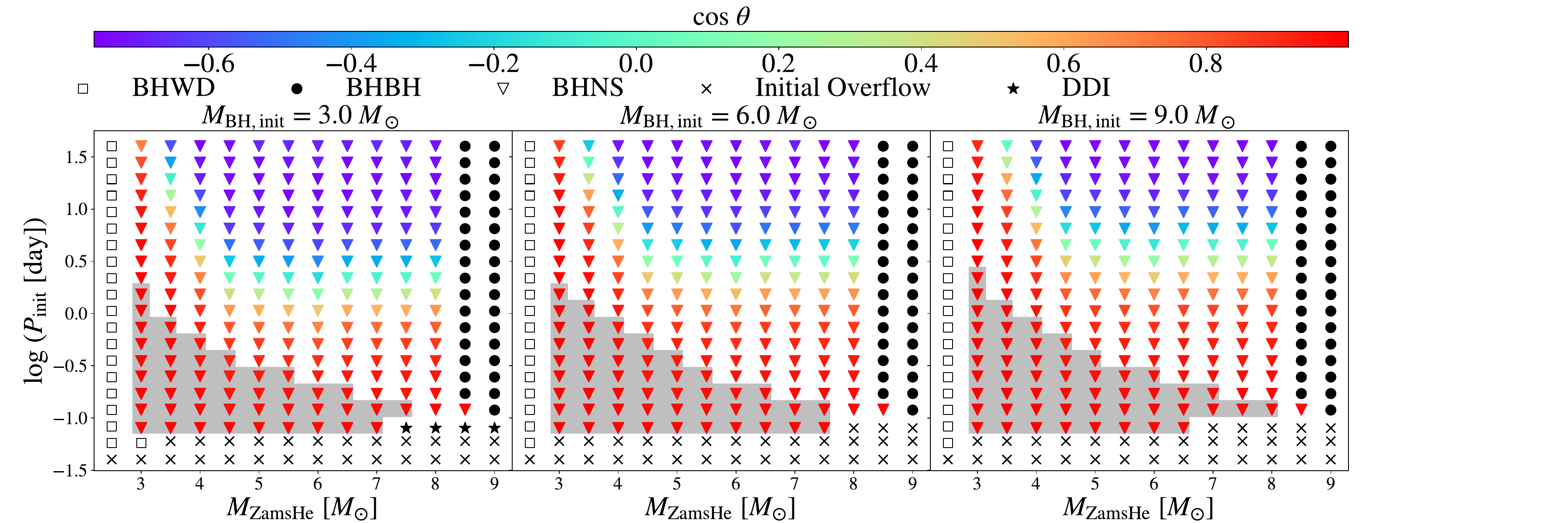}
     \caption{As in Figure \ref{fig5}, but the color bar refer to  $\cos\theta$ values given in Equation (\ref{theta}). The grey region is marked for the parameter space in which Case BB/BC MT occurs when forming BHNS binaries.}
     \label{fig6}
\end{figure*}

Whether an NS can be tidally disrupted by a BH is exclusively determined by the amount of the remnant mass after the BHNS merger (hereafter abbreviated as remnant mass), which can be given by a fitting formula from \cite{foucart2018}, i.e.,

\begin{equation}
\label{equ:fittingMass}
    \frac{M_{\rm total,fit}}{M^{\rm b}_{\rm NS}} = \left[{\rm max}\left(\alpha\frac{1-2C_{\rm NS}}{\eta^{1/3}}-\beta \tilde{R}_{\rm ISCO}\frac{C_{\rm NS}}{\eta}+\gamma\right),0\right]^\delta,
\end{equation}
where $\alpha=0.406$, $\beta=0.139$, $\gamma=0.255$, $\delta=1.761$ are the fitting parameters obtained following numerical relativity simulations, $\tilde{R}_{\rm ISCO}=3+Z_2-{\rm sign}(\chi_{{\rm BH},z})\sqrt{(3-Z_1)(3+Z_1+2Z_2)}$ is the normalized innermost stable circular orbit radius \citep{Bardeen1972} along the orbital angular momentum, with $Z_1=1+(1-\chi_{{\rm BH},z}^2)^{1/3}\left[(1+\chi_{{\rm BH},z})^{1/3}+(1-\chi_{{\rm BH},z})^{1/3})\right]$, $Z_2=\sqrt{3\chi_{{\rm BH},z}^2+Z_1^2}$, $\eta=Q/(1+Q)^2$, and $Q=M_{\rm BH}/M_{\rm NS}$ is the mass ratio. $M_{\rm NS}^{\rm b}$ is the baryonic mass of the NS, which can be expressed as an empirical formula with the NS mass for different EoS introduced by \cite{Gao2020}:

\begin{equation}
\label{equ:baryonicMass}
    M^{\rm b}_{\rm NS} = M_{\rm NS} + A_1M_{\rm NS}^2 + A_2M_{\rm NS}^3,
\end{equation}
where $M^{\rm b}_{\rm NS}$ and $M_{\rm NS}$ are in units of $M_\odot$ . We use the commonly employed EoS DD2 \citep{Typel2010}, with a Tolman-Oppenheimer-Volkoff mass of $2.42\,M_\odot$, which is close to the maximum NS mass we have defined for our detailed binary evolution simulations (see Section \ref{sec:MESAModel}), as a representative EoS. In Equation (\ref{equ:baryonicMass}), $A_1=0.046$, and $A_2=0.010$ for DD2 given by \cite{Gao2020}. We calculate the NS compactness $C_{\rm NS}$ following \cite{coughlin2017}:

\begin{equation}
    C_{\rm NS} = 1.1056\left(\frac{M^{\rm b}_{\rm NS}}{M_{\rm NS}}-1\right)^{0.8277}.
\end{equation}
BHNS merger is considered to be a plunging event if $M_{\rm total,fit}=0$, while the NS can be tidally disrupted by the BH companion to generate bright EM signals when $M_{\rm tot,fit}>0$. We note that the remnant mass is not equivalent to the ejecta mass. However, a higher remnant mass implies that the EM signals from BHNS mergers can be bright. Hereafter, we use the remnant mass to scale the brightness of the EM counterparts.

In Section \ref{sec:BHNSResult}, we explore the intrinsic spin of BH in BHNS binaries by considering different accretion limits during the Case BB/BC MT phase. Nevertheless, whether the tidal disruption of BHNS mergers can occur depends on the projected aligned spin of the BH, as indicated by Equation (\ref{equ:fittingMass}), rather than its intrinsic spin. The kick may result in a tilt angle $\theta$ between the angular momentum of the BH spin and the orbital plane, which could affect the tidal disruption of BHNS mergers. Following Equation (\ref{theta}), we present the averaged $\cos\theta$ value for each BHNS system in Figure \ref{fig6}. It is shown that the tilt angle tends to be larger when the He-rich star is initially more massive and the initial orbit is wider, potentially producing a stronger kick and a wider pre-SN orbital separation. This result is consistent with the earlier finding \cite[see their Figure 1 in][]{zhuxj2021}. On the one hand, as shown in Figure \ref{fig6}, we find that the kick has little effect on the tilt angle of binary systems that experience Case BB/BC MT to spin up BH components. On the other hand, the progenitor systems with $M_{\rm ZamsHe} \gtrsim 3.5 \,M_\odot$ and $P_{\rm init} \gtrsim 2-3\,{\rm days}$ can finally remain BHNS binaries consisting a negative-spin BH. However, due to the weak tidal effect to inefficiently accelerate the BH progenitors \footnote{Figure 2 in \cite{Qin2018} shows that tides can be efficient for binary systems with $P_{\rm init} \lesssim 2.0 \,{\rm days}$.} and the absence of an MT process to spin up the BHs, these BHs thus typically have near-zero spins. Therefore, one can directly use the intrinsic BH spin in Figure \ref{fig4} to judge whether or not the tidal disruption happens for a GW BHNS merger.

\begin{figure}
     \centering
     \includegraphics[width=1.0\linewidth,trim = 20 25 7 45, clip]{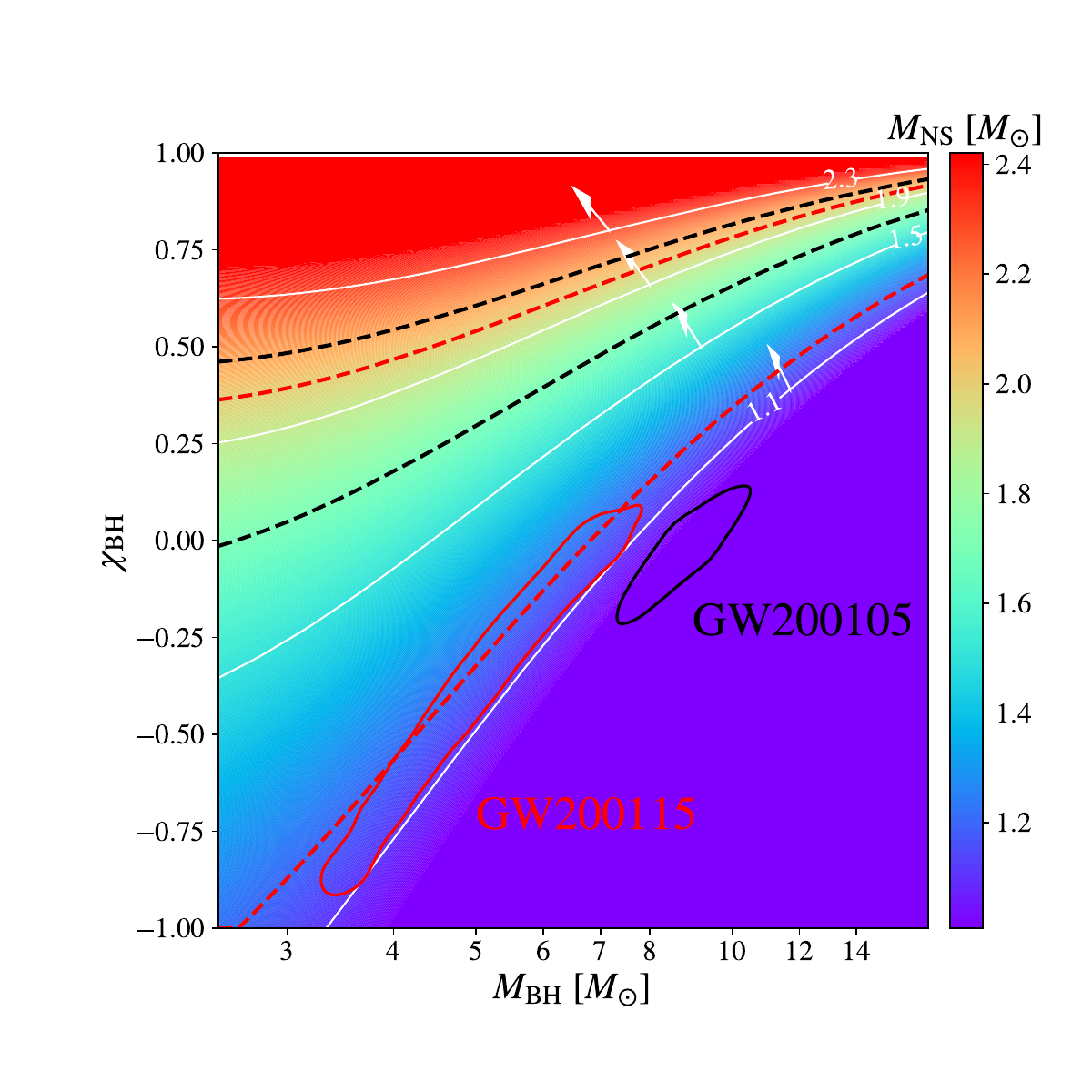} 
     \caption{Parameter space (represented by the direction of the white arrows for specific NS mass) of tidal disruption as a function of BH projected aligned spin and its mass with different masses of NS (the color bar). NS mass contours for $1.1,\,1.5,\,1.9,$ and $2.3\, M_\odot$ are marked with solid white lines. For GW200105 (black solid line) and GW200115 (red solid line), the 90\% credible posterior distributions of the parameters obtained from \cite{abbott2021observation} are displayed, respectively. Corresponding 90\% confidence interval of $M_{\rm NS}$ for these two sources are marked by dashed lines (red dashed lines: GW200115, black dashed lines: GW200105).} 
     \label{fig8}
\end{figure} 

\begin{figure*}
     \centering
     \includegraphics[width=1.0\textwidth, trim = 25 88 30 65, clip]{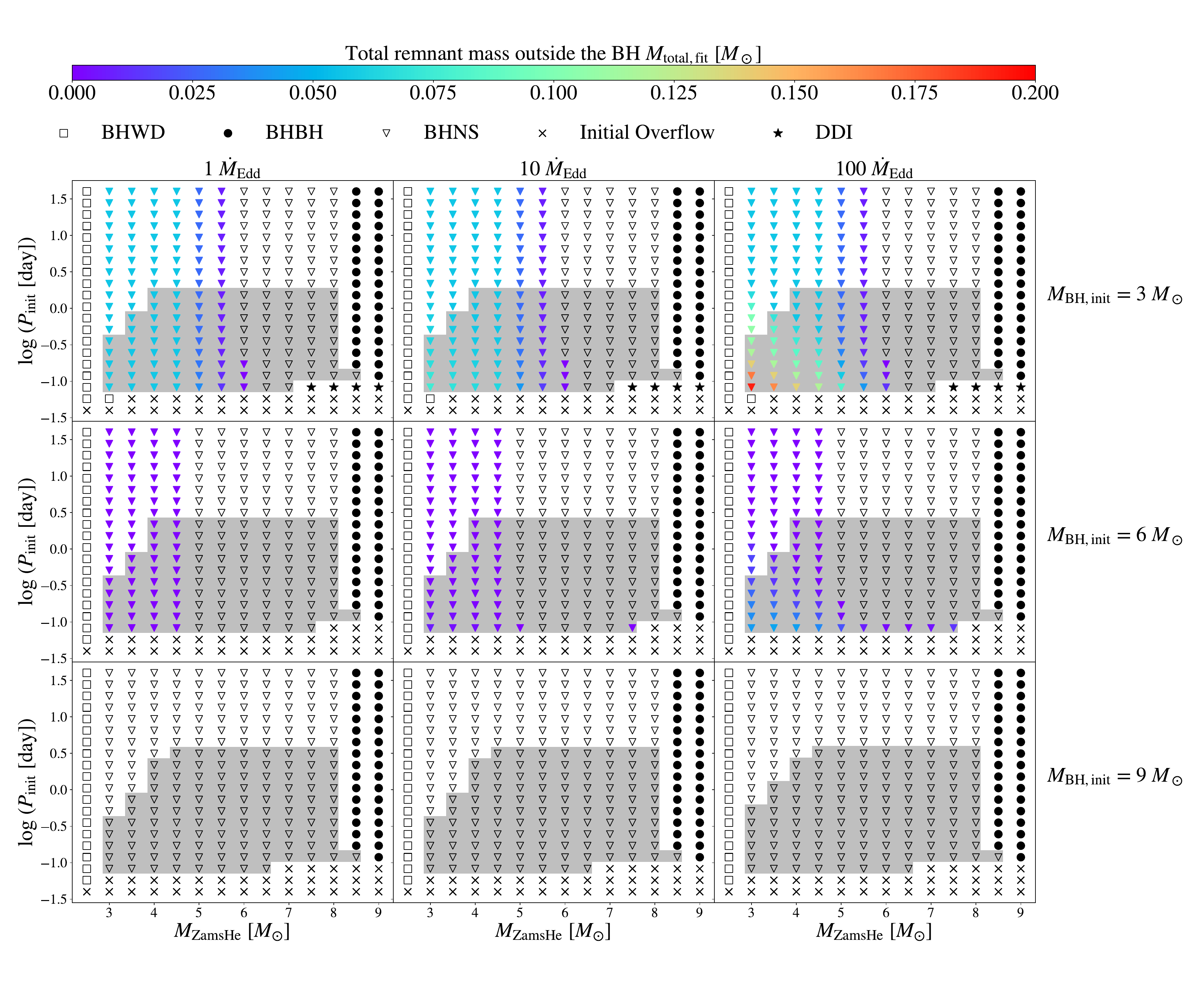}
     \caption{As in Figure \ref{fig5}, but the color bar refers to the total remnant mass outside the BH. The grey background masks the parameter space in which the merger time of BHNS binaries due to the GW emission is not longer than Hubble time.}
     \label{fig7}
\end{figure*}

Following Equation (\ref{equ:baryonicMass}), Figure \ref{fig8} illustrates the parameter space in which NSs can be tidally disrupted by BHs. Tidal disruptions are more likely to occur for binary systems comprising a lower-mass BH with higher projected aligned spin and a lower-mass NS. Given the posterior distributions of the parameters of GW200105 and GW200115 inferred by the LVK collaboration \citep{abbott2021observation}, our detailed calculations could reproduce the two merging events. It also shows that the two systems are most likely plunging events as shown in Figure \ref{fig8}, consistent with earlier findings \cite[i.e.,][]{zhu2021no,broekgaarden2021formation,fragione2021NSBH,Hu2022}. On the one hand, despite BHs having near-zero spins, BHNS mergers involving a representative Galactic NS with a mass of $M_{\rm NS} = 1.4\,M_\odot$ can easily become disrupted events if the mass of the BH component is lower, i.e., $M_{\rm BH}\lesssim 6\,M_\odot$. On the other hand, as the BH mass increases, the required minimum spin magnitude allowing for tidal disruptions in BHNS mergers also increases accordingly. When considering BHs with a mass of $M_{\rm BH}\gtrsim9\,M_\odot$, BHNS mergers can always result in plunging events unless the BH spin is $\chi_{\rm BH}\lesssim0.5$, which are hard to be reached even at high accretion limits (see Figure \ref{fig4}). Therefore, we expect that accretion-induced spin-up would have an insignificant impact on the tidal disruption probability for BHNS mergers. Similar findings can also be obtained from Figure \ref{fig7}, where we evaluate tidal disruptions under various initial conditions and different accretion limits.

\subsection{Remnant mass of disrupted BHNS mergers}
In Figure \ref{fig7}, we estimate the remnant mass for disrupted BHNS mergers with consideration of various initial conditions and different accretion limits. BHNS binaries, capable of merging within the Hubble time, are denoted by the grey background. When the BH accretion rate is constrained to the standard Eddington rate (i.e., $1\dot{M}_{\rm Edd}$), most primary BHs cannot be efficiently spun up through accretion (see Figure \ref{fig4}). The initial orbital period does not show any impact on the amount of the remnant mass. A more massive He-rich star tends to form a higher-mass NS, allowing for the ejection of less remnant mass when the NS merges with its BH companion. As a higher accretion limit is allowed, lower-mass BHs in close-orbit BHNS binaries with smaller $P_{\rm init}$ can be more significantly spun up through accretion (refer to Figure \ref{fig4}), thus easily ejecting more materials when BHNS mergers occur. This trend is more prominent for high accretion limits, especially under a rate of $100\,\dot{M}_{\rm Edd}$. Therefore, it is expected that BHNS binaries with a higher accretion rate limit can eject more remnant mass and thus generate brighter EM signals upon merging. It is worth noting that accretion-induced spin-up is less efficient when compared with BHs spun up through tidal interaction on progenitors (see \citetalias{Hu2022}).

Based on Figure \ref{fig7}, we can further explore the effect of different accretion limits on population properties of the tidal disruption and remnant mass. To achieve this, we perform multi-dimensional interpolation using the calculated remnant mass in Figure \ref{fig7}. In what follows, we describe how the initial parameter distribution of BH--He-rich star binary systems just after the CE phase is adopted. Referring to Figure 14 in \cite{broekgaarden2021impact}, the distribution of initial BH mass in BH--He-rich star binaries is highly dependent on the model assumptions. For simplicity, we adopt three representative empirical BH mass spectra in the literature (see the first column in Table \ref{tab1}), including I) Exponential decay given by \cite{ozel2010}:
\begin{equation}
    P(M_{\rm BH})=
     \begin{cases}
    {\rm exp}[(M_{\rm c}-M)/M_{\rm scale}]/M_{\rm scale}&, M > M_{\rm c}\\
    0&, M\leqslant M_{\rm c}
    \end{cases}
\end{equation}
where $M_{\rm c}=2.5\,M_\odot$, $M_{\rm scale}=1.57\,M_\odot$ (see also \citealp{Steiner2012}).  II) Uniform BH mass distribution in a range of 2.5 -- 5.0 $M_\odot$ + Power law function from 5.0 $M_\odot$ to 15 $M_\odot$ with a slope index of --4.8, and here we adopt 5.0 $M_\odot$ as a connecting point between the two functions \cite[also see Power law distribution of BH mass in][]{Farr2011,Abbott2016}; III) Normal distribution: $P(M_{\rm BH}/M_\odot)\sim \mathcal{N}(\mu = 5.0, \sigma=2.5)$ \cite[the same BH mass function as in][but with different values of mean and standard deviation in order to match the observation of GWTC-3]{ozel2010}. In addition, we simply assume a Power law distribution for the initial He-rich star mass \cite[i.e., $dN/dM_{\rm ZamsHe}\varpropto M_{\rm ZamsHe}^{-3.3}$, similar to the initial mass function as in ][but with an index of -3.3]{Kroupa2001} within a range of [2.5,10]\,$M_\odot$ in which NS is formed, and adopt a uniform distribution for the initial periods of BH--He-rich binaries in logarithmic space, i.e., [0.04, 40]\,$\rm days$ \cite[similar to the range as in][see the Section 4.3]{zhang2023}.

\begin{table*}[]
    \centering
    \renewcommand\arraystretch{1.25}
    \begin{tabular}{lccc}
    \toprule 
    \quad\quad BH mass spectrum& $M_{\rm total,fit}$ [$10^{-2}$\,$M_{\odot}$] ($1\,\dot{M}_{\rm Edd}$)& $M_{\rm total,fit}$ [$10^{-2}$\,$M_{\odot}$] ($10\,\dot{M}_{\rm Edd}$)& $M_{\rm total,fit}$ [$10^{-2}$\,$M_{\odot}$] ($100\,\dot{M}_{\rm Edd}$)\\
    \midrule  
    Model I: Exponential decay& $3.89^{+3.65}_{-3.86}$& $4.08^{+4.27}_{-4.05}$& $6.0^{+11.6}_{-5.9}$\\
    Model II: Uniform + Power law& $3.74^{+3.61}_{-3.60}$& $4.08^{+4.02}_{-3.94}$& $6.5^{+10.3}_{-6.2}$\\
    Model III: Normal & $0.85^{+5.99}_{-0.84}$& $0.90^{+6.40}_{-0.88}$& $2.6^{+11.1}_{-2.6}$\\
    \bottomrule 
    \end{tabular}
    \caption{Median values with 90\% credible intervals of remnant mass $M_{\rm total,fit}$ for BHNS mergers with different assumptions of BH mass spectrum. \textit{Column} 1: BH mass spectrum. The \textit{Column} 2, 3, and 4 correspond to the remnant mass for 1\,$\dot{M}_{\rm Edd}$ 10\,$\dot{M}_{\rm Edd}$ and 100\,$\dot{M}_{\rm Edd}$, respectively.}
    \label{tab1}
\end{table*}

Based on different parameter distributions of the BH--He-rich star binaries, we then perform $10^5$ sampling calculations. First of all, our findings show that the probabilities of the tidal disruption of BHNS systems are moderately dependent on the assumption of BH mass spectrum, i.e., Model I: 0.799, Model II: 0.841, and Model III: 0.702 for the standard Eddington accretion limit. Furthermore, given the same assumption of BH mass spectrum, the corresponding probabilities of the tidal disruption slightly increase with higher Eddington accretion limits, i.e., 10\,$\dot{M}_{\rm Edd}$ (100\,$\dot{M}_{\rm Edd}$): 0.808 (0.818), 0.842 (0.855), and 0.705 (0.736). Thus, a higher accretion-rate MT does not significantly lead to more BHNS mergers undergoing tidal disruption, consistent with the discussion above in Figures \ref{fig8} and \ref{fig7}.  Notably, the fraction of disrupted events in BHNS mergers was recently inferred to be $0.6-9\%$ in \cite{drozda2020}, $0-18.6\%$ in \cite{Xing2023}, and $\sim0-70\%$ in \cite{broekgaarden2021impact}, which are highly determined by model assumptions. Our simulated fraction of disrupted events is slightly higher than that in these studies. This difference could be attributed to the BH spectra used in our work, which tend to generate more low-mass BHs with masses of $\lesssim7\,M_\odot$.

\begin{figure}
     \centering
     \includegraphics[width=0.5\textwidth]{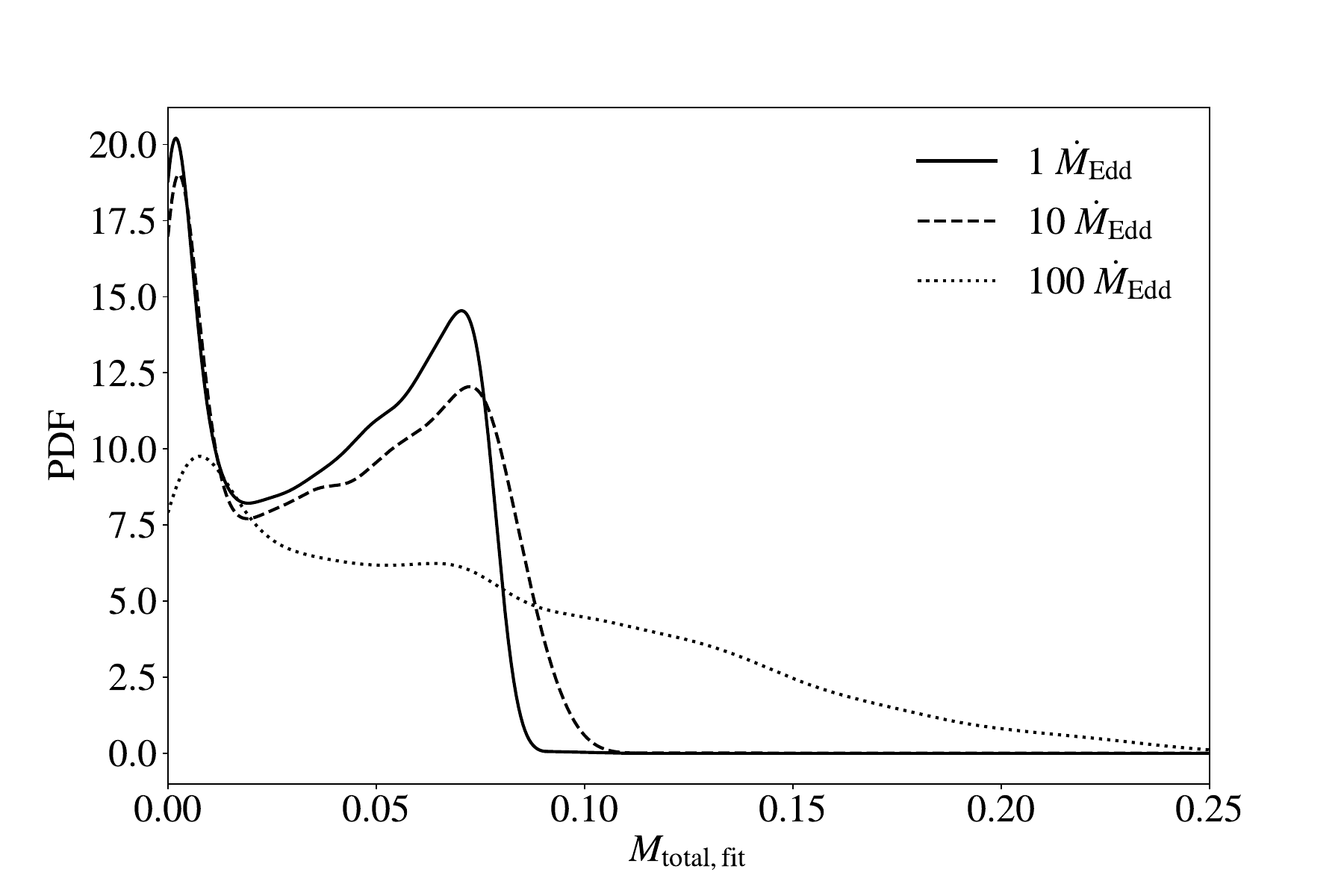}
     \caption{Probability density function of remnant mass $M_{\rm total,fit}$ in BHNS binaries for example Model I in Table \ref{tab1}. Solid, dashed, and dotted lines represent the accretion limits of 1\,$\dot{M}_{\rm Edd}$, 10\,$\dot{M}_{\rm Edd}$, and 100\,$\dot{M}_{\rm Edd}$, respectively.} 
     \label{fig9}
\end{figure}

In Table \ref{tab1}, by assuming various Eddington accretion limits, we present the median value with a 90\% credible level of the remnant mass $M_{\rm total,fit}$ for disrupted BHNS population. We note that the BH mass spectrum plays a key role in determining the remnant mass. However, across different BH spectrum models, the trend of changes in the distribution of the remnant masses remains consistent with varying accretion limits. The amount of the remnant mass when adopting $1 \dot{M}_{\rm Edd}$ shows a slight increase when compared to that for $10 \dot{M}_{\rm Edd}$. However, if the accretion rate is limited to $100\,\dot{M}_{\rm Edd}$, the median and upper limit of the remnant mass increase significantly. As an example, we present the probability density function (PDF) of the remnant mass for Model I in Figure \ref{fig9}. Compared to $1 \dot{M}_{\rm Edd}$ and $10 \dot{M}_{\rm Edd}$, where the remnant mass is always below $\sim0.1\,M_\odot$, the PDF obtained using $100 \dot{M}_{\rm Edd}$ can have a much broader distribution with a maximum value even extending up to $\sim0.25\,M_\odot$. Thus, we can conclude that the population properties of disrupted BHNS mergers by considering different accretion limits also reveal that higher Eddington accretion limits show an insignificant effect on the tidal disruption probabilities. However, they can contribute to the generation of more bright EM counterparts from the BHNS population. 

\section{Conclusions and Discussion} \label{sect4}
As presented in \citetalias{Hu2022}, we study a channel (channel I) to form BHNS binaries in which the tides in close orbits can efficiently spin up the progenitor of BHs and thus produce fast-spinning BHs \cite[also see earlier results in][]{Qin2018,Bavera2020}. In this work, we explore an alternative channel (channel II) where the BH forms first to produce fast-spinning BHs in BHNS binaries as potential GW sources. It is expected the relative rate density of BHNS binaries formed through channel II is higher when comparing with BHNS binaries formed through channel I. Assuming the first-born BH with a near-zero spin \citep[e.g., $\chi_{\rm BH} \sim 0.01$, predicted by the revised version of Tayler-Spruit dynamo as in][]{Fuller2019}, we first perform detailed binary modeling for the evolution of the immediate progenitors of BHNS binaries, in which BHs and massive He-rich stars are in close orbits.

For the initial parameter space, we cover the mass of the He-rich star from 2.5\,$M_\odot$ to 9.0\,$M_\odot$ and the BH companion with a mass range of $3.0-15.0\,M_{\odot}$, as well as the orbital period in $0.04 - 40\,{\rm days}$. In order to investigate how the BH can be significantly spun up through accretion, we assume three different Eddington accretion limits, i.e., 1\,$\Dot{M}_{\rm Edd}$, 10\,$\Dot{M}_{\rm Edd}$, and 100\,$\Dot{M}_{\rm Edd}$. In the standard Eddington accretion limit, the mass accreted onto BHs via Case BB (Case BC) MT can reach $\sim0.01\,M_\odot$ ($\sim0.01\, M_\odot$). The amount of accreted mass is rarely affected by the BH companion mass. However, assuming 100\,$\Dot{M}_{\rm Edd}$, a 3\,$M_\odot$ BH as the companion accretes mass larger than 1.0\,$M_\odot$ via Case BB MT. In this case, the BH can be efficiently spun up (e.g., $\chi_{\rm BH} > 0.6$). We note that the fraction of binary systems that experience Case BB MT is $\sim 9\%$, some of which could potentially lead to fast-spinning BHNS binaries with super-Eddington accretion. When considering SN kicks imparted onto the newly formed NS, we note that the initial orbital period $P_{\rm init}$ for producing mergeable BHNS binaries (i.e., $T_{\rm merger}$ $<$ Hubble time) is required to be shorter than $\sim 2\,{\rm days}$. Additionally, we find the SN kicks play a negligible role in tilting the plane of the newly-formed BHNS binaries when $M_{\rm ZamsHe} \lesssim 3.5\, M_\odot$ and $P_{\rm init} \lesssim 2-3\,{\rm days}$, resulting in no impact on the amount of remnant mass of mergeable BHNS systems within Hubble time. 

Based on the detailed binary modeling of our intense \texttt{MESA} grids, we further adopt multi-dimensional interpolation to estimate the remnant mass during the tidal disruption of the BHNS binaries. Assuming various parameter distributions of BH--He-rich star binaries (component masses and orbital period) to study the population properties of BHNS mergers, we find that the BH mass spectrum plays an essential role in determining the amount of remnant mass of the tidal disruption in BHNS binaries. Based on our simulated population properties of disrupted BHNS mergers by considering different accretion limits, our results reveal that higher accretion limits show an insignificant effect on the tidal disruption probabilities. However, they can contribute to the generation of more bright EM counterparts from the BHNS population. 

The observability of EM counterparts of BHNS mergers is dependent mainly on the magnitude of the BH-aligned spin. The physical process of forming fast-spinning BHs in BHNS binaries remains uncertain. On the one hand, if NSs form first, the BH progenitors can be efficiently spun up by strong tides, resulting in fast-spinning BHs \cite[e.g.,][]{Qin2018,Bavera2020,Hu2022}. On the other hand, the first-born BH could inherit a very high spin from its progenitor assuming a weak angular momentum transport between the stellar core and its envelope \citep{Qin2019,Qin2022AM}. The physics of the angular momentum transport within stars is still uncertain. Angular momentum is transported mainly by meridional currents, the stellar core-envelope coupling is less efficient \citep[e.g.,][]{Georgy2012}, which allows the core to retain more angular momentum and thus forms fast-spinning BHs. 

It is worth noting that some other physical processes could have an impact on the spin magnitudes of resulting BHs. We highlight that the magnetic braking \citep[i.e.,][]{Mestel1968} is not taken into account when modeling low-mass He-rich stars. Therefore, He-rich stars could potentially experience spin-down arising from the coupling of the stellar wind to a magnetic field.

As shown in \cite{Xing2023}, BHs are found to be always formed first and their spins are typically low (i.e., $\chi_{\rm BH} \lesssim 0.2$). However, some BHs can be subsequently spun moderately up ($X_{\rm BH}\, \sim 0.4$) through stable Case BB MT. We note that the inferred probabilities of tidal disruption for BHNS binaries from different groups are very different \citep{broekgaarden2021impact,Xing2023}. These findings mainly arise from the specific parameter distributions of BHNS binaries. For instance, the BH mass spectrum, which depends on the model assumptions, is highly uncertain. We expect that the LVK Collaboration will detect more BHNS binaries in upcoming observing runs (O4 and O5), which will help uncover the properties of BHNS mergers.

\acknowledgments
This work was supported by Anhui Provincial Natural Science Foundation (grant No. 2308085MA29) and the Natural Science Foundation of Universities in Anhui Province (grant No. KJ2021A0106). YQ acknowledges the support of the Key Laboratory for Relativistic Astrophysics at Guangxi University. JPZ thanks the COMPAS group at Monash University. QWT acknowledges support from the Natural Science Foundation of Jiangxi Province of China (grant No. 20224ACB211001). This work was partially supported by the National Natural Science Foundation of China (grant Nos. 12065017, 12192220, 12192221, 12133003, U2038106)),

The inlists and input files to reproduce our simulations and associated data products are available at \dataset[https://doi.org/10.5281/zenodo.10477135]{https://doi.org/10.5281/zenodo.10477135}. All figures are made with the free Python module Matplotlib \citep{Hunter2007}.

\bibliography{ref}{}
\bibliographystyle{aasjournal}

\end{document}